# Locally embedded presages of global network bursts

Satohiro Tajima[1,2,3,*], Takeshi Mita[4], Douglas J. Bakkum[5], Hirokazu Takahashi[4], and Taro Toyoizumi[3]

1) Department of Neuroscience, University of Geneva, CMU, Rue Michel-Servet 1, Genève, 1211, Switzerland.
2) JST PRESTO, Japan Science and Technology Agency, 4-1-8 Honcho, Kawaguchi, Saitama, 332-0012, Japan.
3) RIKEN Brain Science Institute, 2-1, Hirosawa, Wako, Saitama, 351-0198, Japan.
4) Research Center for Advanced Science and Technology, the University of Tokyo, 4-6-1, Komaba, Meguro-ku, Tokyo, 153-8904, Japan.
5) ETH Zurich, Department of Biosystems Science and Engineering, Basel, Switzerland.

*satohiro.tajima@gmail.com, CMU, Rue Michel-Servet 1, Genève, 1211, Switzerland.

**Abstract**

Spontaneous, synchronous bursting of neural population is a widely observed phenomenon in nervous networks, which is considered important for functions and dysfunctions of the brain. However, how the global synchrony across a large number of neurons emerges from an initially non-bursting network state is not fully understood. In this study, we develop a new state-space reconstruction method combined with high-resolution recordings of cultured neurons. This method extracts deterministic signatures of upcoming global bursts in "local" dynamics of individual neurons during non-bursting periods. We find that local information within a single-cell time series can compare with or even outperform the global mean field activity for predicting future global bursts. Moreover, the inter-cell variability in the burst predictability is found to reflect the network structure realized in the non-bursting periods. These findings demonstrate the deterministic mechanisms underlying the locally concentrated early-warnings of the global state transition in self-organized networks.

## Introduction

The collective firing dynamics of neural population have been related to emergent functions and dysfunctions in the brain. Unraveling the structure of dynamics emerging from the collective behavior in neural networks is an ultimate issue in studies of complex systems (1). The simplest and most profound example of collective dynamics is synchronous burst: the simultaneous occurrence of closely-spaced action potentials ("bursts") across a large number of neurons in the network. Within neural circuits in vivo, the synchronous bursts are believed to serve important roles in information storage and transmission (2, 3) as well as in disease states including epileptic seizures (4–6). Intriguingly, neurons cultured in vitro also display spontaneous synchronous bursts akin to those observed in vivo (7–9), indicating that the collective bursting is a universal behavior of neural networks in a wide variety of preparations.

Classic biophysical models of the synchronized bursts assume broadly diverging synaptic interaction that propagates activity of a single cell to the remaining neural population (4, 10–12). This view has been partially supported by experiments demonstrating that stimulating a single neuron can change the large-scale dynamics of neural populations (13–15). These models and experiments suggest a close link between single neuron activity and the global state of a neural population. On the other hand, the previous stochastic models do not specify *when* a global burst occurs; forecasting a spontaneous occurrence of synchronous burst from the preceding individual neural activities in non-bursting states remains a highly challenging issue, particularly in real nervous systems which feature the heterogeneity of local-global coupling (e.g., non-uniform coupling strength between individual neural activity and the overall firing of the population) (16). Indeed, recent studies demonstrate potentially diverse dynamical mechanisms accompanied by complex but reproducible patterns of activity sequences leading to the synchronous bursts (17, 18), implying the existence of nonlinear deterministic mechanisms around the burst initiations.

The present paper investigates the deterministic aspects of burst initiation. In particular, we question whether individual neuron dynamic in non-bursting periods can predict an upcoming synchronous burst in cultured neural population. To cope with the heterogeneous nature of local-global coupling, we introduce a model-free method to quantify the deterministic relationships among individual neuron traces and the global state, based on a nonlinear state-space reconstruction technique. Using this method, we report that dynamic of only one neuron can robustly predict the upcoming synchronous burst in the neural population, at high signal-to-noise ratio. Surprisingly, the burst predictability with an appropriately-chosen neuron even outperforms that with the mean field activity of neural population, demonstrating that macroscopic fluctuation of neural population is better predicted by the microscopic dynamic of a specific single cell, rather than the macroscopic state itself. This apparently counterintuitive property is explained based on mathematical property termed 'embedding' in nonlinear dynamical systems and the heterogeneous input-output interactions among neurons in the network shaping non-bursting spontaneous dynamics.

## Results

### Spontaneous synchronous bursts in cultured neurons

Rat cortical neurons were grown on high-dense CMOS-based multi-electrode arrays for 12-41 days-in-vitro (MEAs; **Fig. 1a**) (19, 20). The devices allow us to simultaneously stimulate and record from multiple neurons lying on the array surface at a high spatiotemporal resolution (**Methods**). Each recording yielded simultaneous traces of 60–98 cells at a temporal resolution of 2 kHz. **Figure 1b** shows the cell distribution recorded with a representative array preparation. In our preparations of sparse culturing, the cell bodies were well isolated, and the action potential of each cell was identified accurately based on its spatial location and waveform. Spike action-potentials were detected using standard LimAda method (21).

We first analyzed the spontaneous activities of the recorded neural population. During the spontaneous firing, the neurons showed intermittent synchronous bursts, which were interleaved with silent periods (**Fig. 1c**). To describe the macroscopic network dynamics, we defined population activity by the mean firing rate across all recorded neurons (**Fig. 1d**). The synchronous bursting period were defined by time bins in which the population activity exceeded a threshold determined for each preparation, according to the previously established method based on inter-spike interval distributions (20) (**Fig. 1d**, inset). The timing of synchronous burst occurrence did not show any clear periodicity. Duration of bursts varied from about 200 to 800 ms. The different preparations showed distinct complex patterns of collective activity dynamics when visualized with standard dimensionality reduction techniques such as principal component analysis (**Supplementary Fig. S1**).

**Predicting bursts**

To capture the potentially complex temporal structures in neural activities around bursting dynamics, we developed a model-free method that is able to quantify the general relationships between the network and single-element dynamics in high-dimensional nonlinear systems. Because the method features tracing the reconstructed state-space trajectories in the retrograde direction, from now on we term the method as *trace-back embedding* (TBE). The TBE extends the nonlinear forecasting techniques based on state-space reconstruction using the delay-embedding theorems in deterministic dynamical systems (22, 23). Compared to the previous nonlinear forecasting methods (24–26), it has two additional advantages. First, the method focuses particularly on the relationships between a global mean-field variable and local variables. Second, the method quantifies the predictability of a global synchronous burst based on temporally distant dynamics of each local variable. The basic protocol of TBE is summarized below.

**Figure 2** illustrates the TBE protocol (further details are provided in **Methods**). We first reconstruct the dynamics of population activity and each single-neuron activity respectively in delay-coordinate state-spaces (**Fig. 2a, b**):

$$\underline{b}_t^d = (b_t, b_{t-\tau} \ldots, b_{t-(d-1)\tau}), \qquad (1)$$

$$\underline{x}_{i,t}^d = (x_{i,t}, x_{i,t-\tau} \ldots, x_{i,t-(d-1)\tau}), \qquad (2)$$

where $b_t$, and $x_{i,t}$ are the population mean-field and neuron $i$'s activity at time $t$, respectively; $d$ denotes the embedding dimensions (number of delay-coordinates); $\tau$ is the unit delay size. In this study, the local time series $x_{i,t}$ was defined by each neuron $i$'s spike train smoothed with a Gaussian kernel, and $b_t$ was defined by the mean of $x_{i,t}$ over all the simultaneously recorded neurons (**Methods**). For convenience, we refer to $\underline{b}_t^d$ and $\underline{x}_{i,t}^d$ as the "global" and "local" state trajectories, respectively. The TBE method measures the accuracy of synchronous-burst prediction by seeking "presages" of global bursts in each local state trajectories, based on the accuracy of forecasting using a nearest-neighbor model (**Fig. 2b**, see also the figure legend for a step-by-step description of the protocol). The rationale behind the TBE protocol is as follows: if neuron $i$ has enough information to predict the future synchronous burst occurring after a time span $-\Delta t$ ($-\Delta t > 0$), neuron $i$'s state up to time $-\Delta t$ before the targeted burst should be already differentiated from the states without any future bursting, forming a cluster of states that deviate from those predicting non-bursting periods in the reconstructed state space. Similarly, we define the accuracy of "postdiction" of bursts (i.e., detecting a burst event based on neural activities *after* its occurrence) with the same protocol as that for prediction except for using time span $\Delta t$ with a positive value. We iterated this procedure for all the neurons to characterize the burst predictabilities in individual neurons. In addition, we quantified the "self-predictability" of the mean-field activity by replicating the above procedure based on its own (global) state trajectory, $b_t$, instead of a local state trajectory in individual neurons, $x_{i,t}$.

Using this method, we investigated the predictability of synchronous bursts based on the individual neural activities or the population mean-field. In particular, we addressed two key questions: (i) how accurately can each single neuron predict the future synchronous bursting events, and (ii) if a subset of neurons predicts synchronous bursts better than others, how are they related to the underlying neural network structure?

**Single-neuron dynamics can predict global network bursts**

First, we asked how accurately individual neurons predict (or postdict) the occurrence of synchronous bursts, and how their predictions compare to that based on the population mean-field, which defines synchronous burst. We quantified the rate of successful synchronous-burst predictions in each single neuron with keeping the false-positive rate at 5% (**Methods**). Interestingly, in some cells (e.g., **Fig. 3a**), the rates of successful burst predictions with the TBE method could be relatively high (>50% of trials) even when the network appeared to be silent in terms of the global mean-field firing rate (**Fig. 3d**, –1000 ~ –500ms). Corresponding to this, the success rate of TBE-based burst detection using a single neuron dynamics could be even higher than that with the global mean field dynamics (**Fig. 3b**). The predictability of burst occurrence varied among neurons, suggesting the heterogeneity of burst predictability across neurons. However, the accuracy of predictions in individual neurons were highly consistent among different sets of burst trials within each preparation ($\rho=0.79$~$0.93$, $P<2\times10^{-23}$, Spearman's rank correlation between the first and the second halves of trials; **Fig. 3e**), demonstrating that this cell-to-cell variability reflects a robust property of each neuron. Across all the preparations, about 1/3 of neurons were found to predict synchronous bursts robustly better than population mean-field over some prediction span, $\Delta t$ (**Fig. 3g**).

The fact that only a single neuron can compare or even outperform the population mean at predicting the mean-field dynamics may appear surprising, considering that the single-cell activities fluctuate due to spiking much more than the mean-field activity (e.g., compare the black traces in **Figs. 3c** and **3d**). However, it is explained by the delay-embedding theorems (22, 23) that the global state information could be reconstructed from observation of time series in a single variable participating in the dynamical system (**Supplementary Fig. S2**). The present result suggests that such a mathematical property can be demonstrated in a biological system. Importantly, to reconstruct the global state requires, we need to observe multiple time points rather than the momentary state in the time-series of the observed variable (which is why the theorem is called "delay-embedding" theorem). It suggests that the information about the global state is carried by the temporal patterns including multiple time points, rather than momentary snapshots of activity in neurons.

To explore the importance of temporal patterns of activities in predicting bursts compared to the momentary firing rates, we compared the current TBE-based prediction accuracy with an analogous index based on momentary firing rates, $x_{i,t}$ or $b_t$ (the red traces in **Figs. 3c** and **3d**, respectively; see also **Methods**). In single neurons, the TBE method tended to provide more accurate prediction than the momentary-activity-based method ($P<5\times10^{-14}$ in all the individual preparations, sign test with paired samples; **Fig. 3f**). This superiority of TBE to the momentary firing rate is not likely to be due to the difference in the sampled spike numbers because it remained even if we lengthened the bin size for computing activity so as to have the same total duration as used in TBE ($P<6\times10^{-3}$ in all the preparations except for Chip 2427 [which had much lower firing rate than others], sign test with paired samples; **Fig. 3f**, the inset). Finally, when we used the momentary firing rates instead of the TBE with temporal sequence, the fraction of cells whose burst detectability outperformed that of the mean field reduced from ~1/3 to ~1/5 (**Fig. 3g**). Together, these results suggest that the temporal patterns in neural activity, not only momentary activity, are crucial for early detection of synchronous bursts.

**Burst predictability reflects network structures realized in non-bursting periods**

Next, we examined whether the predictability of synchronous bursts corresponds to any specific network structure defined by synaptic interactions. The structural correspondences of burst predictability were investigated in three steps (**Figs. 4**): we first re-analyzed the non-bursting spontaneous activities to infer the directed causal network structure; next, the inferred causal networks were verified by comparing them to synaptic interactions measured by an additional electrical stimulation experiment with the same preparations; finally, we related each neuron's network connectivity with its burst prediction accuracy.

In the causal network analysis, we applied a previously proposed method that is capable of detecting weak nonlinear coupling (26) (**Methods**). Theoretically, the method detects causal interactions including indirect ones (27). Applying this method to all the neuron pairs showing non-bursting spontaneous activities yielded a matrix for each preparation that represents pairwise, directed interactions within each preparation (**Fig. 4a**).

The relevance of the inferred causal couplings to synaptic interactions among neurons was verified using independent data in which synaptic interactions were directly measured (**Fig. 4a, b**). The present CMOS-based MEA system allowed us to stimulate the neural tissue local to given electrodes at sub-millisecond accuracy, directly activating a single neuron's soma at a high temporal precision with few artifact (19). A subset of neurons was electrically stimulated while the evoked activities in other neurons were monitored simultaneously. According to the evoked response latencies, we identified the short-latency ($< 10$ ms) and long-latency ($\geq 10$ ms) downstream cells for each stimulated neuron, between which the latter was expected to include more multi-synaptic interactions and thus weaker causal effects. As expected, we confirmed that the causality inferred based on non-bursting spontaneous data consistently reflected the identified short- and long-latency interactions (**Fig. 4b**). Specifically, the inferred causality differentiated the short- and long-latency downstream cells from the remaining population ($P = 0.0001$, sign test, data pooled across the all preparations). Based on these confirmations, we used the inferred causality as proxies for the effective synaptic interactions in the network during the non-bursting periods.

Finally, we compared these causal network structures and the burst predictabilities, finding that the burst predictability in each neuron displayed a positive correlation to the average inferred causality between it and the other neurons (**Figs. 4d and 4e**). We classified putative excitatory and inhibitory neurons based on the spike waveforms (**Methods**). Interestingly, the excitatory neurons predicted bursts better than the inhibitory neurons when we compared the neurons showing strong causality ($P<0.002$, both for the causality for input and output connections, rank sum test), but such cell-type dependency was unclear in the neurons having weak interactions ($P>0.5$; **Figs. 4d**). These results suggest that the stronger excitatory network interaction with other cells underlie the higher burst predictability. The above results together led us to conclude that the cell-to-cell variability in burst predictability is likely to reflect the heterogeneous network structures that shape non-bursting activity dynamics.

**Discussion**

In this study, we have focused on the relationship between local and global neural properties by analyzing how single cell activity predicts the population mean-field dynamics. The newly introduced TBE method revealed the heterogeneity of neurons also in the predictability of upcoming occurrence of a synchronous burst of a neural population, which is temporally distant in order of hundred milliseconds. The bursts were predicted accurately by dynamics of some single neurons. Remarkably, such neuron even outperformed the population mean-field dynamics itself in terms of prediction accuracy, particularly when we extract the information from the temporal patterns of neural activity rather than the momentary firing rates. From causal network analysis combined with the electrical stimulation experiment, the heterogeneity of burst predictability was explained by the structures of neural networks: the "hub"-like neurons having stronger synaptic interactions with other cells can predict the upcoming global network burst better than others. In addition, the present results based on the cultured neurons indicate the heterogeneous relationships between single neurons and population activity are basic properties observed in the nervous network that is self-organized without external stimulus.

Classic models of the synchronized population burst assume diverging interaction that propagates activity of a cell to the other cells (4, 10–12). A variant of this model features "avalanche" like bursting sequences, which models synchronous burst with noise being amplified and propagated through the network (9, 18, 28–32). These models, focusing mainly on the stochastic aspects of burst initiation and propagation, have made successes in explaining a variety of statistical properties of bursting neural networks. On the other hand, it has been challenging to predict the occurrence of synchronous burst, partially because they can be triggered by complex and heterogeneous mechanisms (33). The present study, focusing on deterministic aspects of the neural population dynamics that transit from silent state to synchronous burst, demonstrates that the burst occurrence is predictable to the extent based on preceding activity traces in single neurons. This is in line with studies suggesting that the spatiotemporal pattern of spontaneous fluctuation in neural population activity is not purely random but restricted to a limited number of configurations (34, 35). The existence of specific neurons predicting the upcoming burst is also consistent with the previous idea of 'leader neurons' that fire at the beginning of bursts (8, 36). Notably, however, the predictor of the global burst is not identical to the increased activities in those single neurons as expected from previous studies (13, 30, 36), because the magnitude of FR alone is not sufficient to explain the highly accurate predictability based on TBE method (**Fig. 3e**). Rather, the earliest presages are likely to be embedded in the dynamical patterns of those neurons.

On the other hand, it should be noted that not everything is predictable in a deterministic manner before the burst occurrence. A most relevant example is the trial-to-trial fluctuation in burst size (the height of peak population rate during a burst). In contrast to the case of the burst detection, the cell-to-cell heterogeneity in the burst-size predictability (**Methods**) was not always consistent ($\rho = -0.289 \sim -0.077$, $P>0.4$, Spearman's rank correlation between the odd and even trials). Neither the global state could predict the size of upcoming burst ($\rho = 0.091 \sim 0.35$, $P>0.1$, Spearman's rank correlation between actual and predicted burst sizes). These facts suggest that the burst size is determined in a different mechanism from that triggering the burst: the former is more likely to follow a stochastic mechanism during the burst (as previously described in probabilistic activity-propagation models (18, 32)) whereas, somewhat counterintuitively, the latter seems to follow more deterministic process.

Finally, the predictability of upcoming burst occurrence based on single-cell dynamics suggests an effective methodology to capture the early warnings of population state transition in a variety of networks, including epileptic seizures in the brain (5, 37), rumor propagation in social networks (38–40), or epidemic spreading in human networks (41, 42). In clinical purposes, for example, accurate prediction and characterization of pathologically synchronized neural firing is an important issue (5, 37). Although nonlinear features of field potential sequence have been proposed as useful markers of epileptic state (37, 43–46), the predictability of global dynamics based on single neuron activity, to the best knowledge of ours, has been not thoroughly studied. In particular, the present study suggests that single neuron could even outperform the direct observation of global state itself in predicting the global state dynamics. The success of the present state-space reconstruction-based method implies a new approach to detect the early warnings of transitions in global network state based on local rather than global features.

## Methods

### Cell culturing

All experimental protocol was approved by the Committee on the Ethics of Animal Experiments at the Research Center for Advanced Science and Technology, the University of Tokyo (Permit Number: RAC130106).

A protocol for cell culturing had been reported previously (19). Briefly, E18 Wistar rat cortices were dissociated using trypsin and mechanical trituration. 20-40k/μL neurons and glia were seeded over an area of ~12 mm$^2$ on top of the CMOS chip. Layers of poly (ethyleneimine) followed by laminin were used to adhere cells. Plating media consisted of Neurobasal-B27 supplemented with 10% horse serum and 0.5 mM GlutaMAX during the first 24 hours. Growth media consisted of DMEM supplemented with 10% horse serum, 0.5 mM GlutaMAX, and 1 mM sodium pyruvate. Experiments were conducted inside an incubator to control of environmental conditions (36°C and 5% $CO_2$).

### CMOS-based recording and stimulation of network activity

Cultured neuron activities were simultaneously recorded with high-density microelectrode array as described before (19, 20). Cortical networks were grown over 11,011-electrode CMOS-based MEAs (47, 48), which provide enough spatial and temporal resolution to detect action potentials from any neuron lying on the array: $1.8 \times 2.0$ mm$^2$ area containing $8.2 \times 5.8$ μm$^2$ electrodes with 17.8 μm pitch, sampled at 20 kHz. Subsets of 126 electrodes can be read-out (and stimulated) at one time, and electrode selection can be re-configured within a few milliseconds. To identify the locations of neurons growing over the array, a sequence of about one hundred recording configurations were scanned across the whole array while recording spontaneous activity. Locations of active somata were identified because action potentials could be usually detected from multiple nearby electrodes. Electrode selection was then re-configured such that a single electrode was assigned to each identified soma, and spontaneous activities were measured simultaneously from all of these cells. The putative neuron types (excitatory or inhibitory) were estimated based on spike shapes (**Supplementary Fig. S3**). We could recode from 93, 47, 98, 92 and 53 neurons in Chip 1437, 1440, 1444, 2427 and 2440, respectively.

Microstimulation-elicited network activities were then investigated to characterize a pairwise synaptic strength between neurons. An adequate stimulating electrode was explored such that a single target neuron was directly activated through axonal stimulation and that the directly evoked spikes were exclusively measured from the target cell. The directly evoked spikes could be easily distinguished from post-synaptic activations because they were very reliable (i.e., 100 spikes elicited out of 100 stimulation trials) and exhibited a small temporal jitter (**Supplementary Fig. S4**). The microstimulation was a single, positive-negative, biphasic pulse with a charge-balanced amplitude of ± 300-900 mV and a duration of 200 μs/phase, and was delivered 100 times every 3 sec.

**Burst detection**

Burst in spontaneous activity was detected modifying a protocol that was previously established by authors' group [20]. The previous study proposes to determine the threshold for burst detection based on the distribution of inter-spike interval (ISI) of consecutive spikes [20]. In the bursting neurons, the ISI distribution typically has a bimodal structure whose valley can be used as an objective criterion for burst detection. In this study, we first computed sequences of population firing rate that were normalized such that it ranges from 0 to 1. We defined this as the sequence of global state, $b_t$, where $t$ is the time. This was to use a common burst detection threshold across different preparations of neuron cultures, which generally vary in terms of the absolute firing rate. We used 5 ms bin for the firing rate computation. We next derived the distributions of inverse firing rate over the bins. This yielded an ISI distributions for each preparation, in which we confirmed that all of them had bimodal shapes (**Fig. 1d**, inset). For each preparation, we selected the smallest ISI providing the valley of distribution as the burst detection threshold. The burst periods were determined by the at least three consecutive sequences of bins that have average ISIs under this threshold, identifying 26 bursts on average for each preparation. We defined the peak burst timing by selecting the center of the bin having the smallest average ISI (i.e., the largest global state) within each burst period.

**Estimation of synaptic connectivity from electrical stimulation**

Synaptic connectivity was estimated based on the evoked responses during electrical stimulation experiment. To reliably stimulate a single neuron, we searched a stimulation site around a target neuron in each MEA that could elicit an action potential exclusively at the target neuron. After the careful selection of stimulation sites, we could evoke the action potentials at almost 100% probability for each single stimulation, with very little jitters in the timings of action potentials in the stimulated neurons. We first aligned the spike raster to the timings of stimulation. We next computed the sequence of firing rate $x_{i,t}$ in a way described above. The firing rates before the stimulation (-500 ms $< t <$ -100 ms) were used to produce the reference probability distribution of each neuron's firing rate, P($x_i$). Next, we computed the probability, P($x_i >$ $x_{i,t}$), for each time bin centered at $t$ in a short post-stimulation period (2.5 ms $< t <$ 15 ms); the responses during 0 ms $< t <$ 2.5 ms were omitted to eliminate the artifacts of electrical stimulation. We defined the smallest time $t$ that satisfies P($x_i >$ $x_{i,t}$) $< 0.01$ in the post-stimulation period as the latency of neuron $i$. The neurons that had at least one time-bin satisfying this condition were defined as "downstream| cells of the stimulated neuron; the neurons that do not have any such time bin were defined as "non-downstream" cells.

According to the evoked response latencies, we identified the short- ($< 10$ ms) and long-latency ($\geq 10$ ms) downstream cells for each stimulated neuron. The multi-synaptic downstream cells for a stimulated neuron comprised relatively small fraction n (17% on average) of the entire population. This indicates that they are subsets of neurons receiving effectively strong input from the stimulated cell via relatively a small number of path length, although the interaction with longer path-length, which was not detected here, could include the larger fraction of the cell population.

**Causal network analysis**

Pairwise causal interaction among neurons was estimated based on a variant of convergent cross-mapping (CCM). CCM was developed by Sugihara et al. [26] as an extension of nonlinear forecasting method of time sequence based on nearest-neighbor models [24, 25, 49]. The method is capable of detecting relatively weak causal interactions in deterministic dynamical systems (which can include some stochastic components) if the system and observations are deterministic, the embedding dimension is sufficiently large, and data size is sufficiently large for the given embedding dimension. A variant of CCM that can be used for systems, in which the variables have heterogeneous time-scales (such as in neural system), was developed by the authors (27).

Suppose that we want to know the interaction from neuron $x$ to another neuron $y$. We first reconstruct the state-space trajectories of each neuron $x$ in the delay-coordinates, $\underline{x}_t^{dmax} = (x_t, x_{t-\tau}, ..., x_{t-(dmax-1)\tau})$, where $dmax$ represents the maximum number of dimensions (number of delay-coordinates) to be considered, $t$ is the time point, and $\tau$ is the unit delay length. We used $\tau = 5$ ms $dmax = 8$ in this study. In the embedding-based analyses, we convolved the firing-rate sequence of each neuron with Gaussian kernel that has half-width-of-half-height of 25 ms, and normalized such that each neuron trace ranges from 0 to 1. To avoid a known vulnerability of the state-space reconstruction to the time-scale heterogeneity (50), we projected delay vector $\underline{x}_t^{dmax}$ to a randomized coordinate space by multiplying a square random matrix, $R$, from the left to obtain a transformed vector: $x_t^{dmax} = R \underline{x}_t^{dmax}$. A $d$-dimensional delay vector $x_t^d$ is constructed by selecting the first $d$ ($\leq dmax$) components of $x_t^{dmax}$. The causal interaction form neuron $x$ to another neuron $y$ is quantified based on the correlation coefficient, $\rho_{y_t, \hat{y}(x_t^d)}$, between the true ($y_t$) and forecast ($\hat{y}_t(x_t^d)$) signals, where

$$\hat{y}_t(x_t^d) = \sum_{t' \text{ s.t. } x_{t'}^d \in B(x_t^d)} w(|x_{t'}^d - x_t^d|) \, y_{t'}$$

with the k-nearest neighbor set $B(x_t^d)$ of $x_t^d$ in the delay-coordinate space. We set $k$=4, weight,

$$w(|x_{t'}^d - x_t^d|) = \frac{\exp(-|x_{t'}^d - x_t^d|)}{\Sigma_{t' \text{ s.t. } x_{t'}^d \in B(x_t^d)} \exp(-|x_{t'}^d - x_t^d|)},$$

and $|x_{t'}^d - x_t^d|$ as the square distance between $x_{t'}^d$ and $x_t^d$ in the data analysis. Note that $\rho_{y_t, \hat{y}(x_t^d)} = 1$ means perfect prediction. We selected the optimal dimension $d=d^*$ so as to maximize the prediction performance, $\rho_{y_t, \hat{y}(x_t^d)}$. Typical value of $d^*$ was distributed from 2 to 4 in the current data.

**Trace-back embedding method and burst predictability**

The trace-back embedding (TBE) method is an extension of the above CCM algorithm. The method extends the original CCM (i) by relating the global state to local components while the original protocol only quantified the relationships between local variables, and (ii) by quantifying on the predictability of temporally distant global state based on local variable dynamics. This is done by finding a mapping between the system's global state $b_t$ and each neuron $x_{t-\Delta t}$ including temporal lag $\Delta t$, instead of simultaneous prediction among individual neurons. For burst detection, we set $t$ as the time corresponding to the peak of global state during a burst period, $t \in \{t_1, ..., t_n\}$, where $n$ is the number of burst trials in the experiment. We first reconstruct the state-space trajectory for neuron $x$ and the normalized global state $b$ in the (randomized) delay-coordinates, $x_t^d = (x_t, x_{t-\tau} ..., x_{t-(d-1)\tau})$ and $b_t^d = (b_t, b_{t-\tau} ..., b_{t-(d-1)\tau})$. It is mathematically shown that both of these trajectories reproduce the topological structure (i.e., preserving continuity but allowing deformations) of attractor dynamics that they participate, with which a (near-) future state of each variable is accurately predicted from the current state of other variables, if the dynamics (approximately) follows deterministic mechanisms (22, 23, 26, 27). Practically, however, the accuracy of prediction is limited by various factors including data size, process/observation noises, as well as the uncertainty due to asymmetric causal interactions.

The forecast of the global state $\hat{b}_t$ based on a single neuron $x$ with time lag $\Delta t$ is given by

$$\hat{b}_t(x_{t-\Delta t}^d) = \sum_{t' \text{ s.t. } x_{t'-\Delta t}^d \in B(x_{t-\Delta t}^d)} w(|x_{t'-\Delta t}^d - x_{t-\Delta t}^d|) \, b_{t'}$$

with the k-nearest neighbor set $B(x_{t-\Delta t}^d)$ of $x_{t-\Delta t}^d$, in the same manner s in CCM. In the data analysis we used $k$=4 and the same weight function,

$$w(|x_{t'-\Delta t}^d - x_{t-\Delta t}^d|) = \frac{\exp(-|x_{t'-\Delta t}^d - x_{t-\Delta t}^d|)}{\Sigma_{x_{t'-\Delta t}^d \in B(x_{t-\Delta t}^d)} \exp(-|x_{t'-\Delta t}^d - x_{t-\Delta t}^d|)}.$$

The success rate in burst detection was defined by the ratio between the number of burst trials such that $\hat{b}_{t_n} > \theta_x$, where $\theta_x$ is the threshold of burst detection for neuron $x$. In the present study, we defined $\theta_x$ by the 95 percentile in the null-hypothesis distribution $P(\hat{b}_{t_{\text{Null}}}(x_{t_{\text{Null}}-\Delta t}^d))$, where $t_{\text{Null}}$ is randomly selected time outside the bursting period. The burst-size predictability in each cell was defined by correlation between actual and predicted burst sizes, average over range of time span -200 ms < $\Delta t$ < -50 ms.

Similarly, the forecast based on the history of the global state itself is given by the same protocol except for using $\boldsymbol{b}$ instead of $\boldsymbol{x}$:

$$\hat{b}_t(\boldsymbol{b}^d_{t-\Delta t}) = \sum_{t' \text{ s.t. } \boldsymbol{b}^d_{t'-\Delta t} \in B(\boldsymbol{b}^d_{t-\Delta t})} w(|\boldsymbol{b}^d_{t'-\Delta t} - \boldsymbol{b}^d_{t-\Delta t}|) \, b_{t'}.$$

This quantifies the self-predictability of burst based on the global state itself, through the success rate compute based on the corresponding threshold $\theta_b$, which in the present study was defined by 95 percentile of null-hypothesis distribution $P(\hat{b}_{t_{\text{Null}}}(\boldsymbol{b}^d_{t_{\text{Null}}-\Delta t}))$.

## Acknowledgements
The authors are grateful to Dr. Urs Frey at RIKEN Quantitative Biology Center, Kobe, Japan, and Prof. Andreas Hierlemann at ETH Zurich, Basel, Switzerland, for providing CMOS-based MEA with their technical supports. This work was supported by JST PRESTO (ST), Asahi Glass Foundation, and KAKENHI (26630089), Brain/MINDS from AMED (TT), and RIKEN Brain Science Institute (TT).

## Conflict of interests
The authors declare no competing financial interests.

## Author contributions
Designed research: ST, TM, HT, TT
Performed research: -
Contributed unpublished reagents/analytic tools: TM, DJB, HT
Analyzed data: ST, TT
Wrote the paper: ST, TM, HT, TT

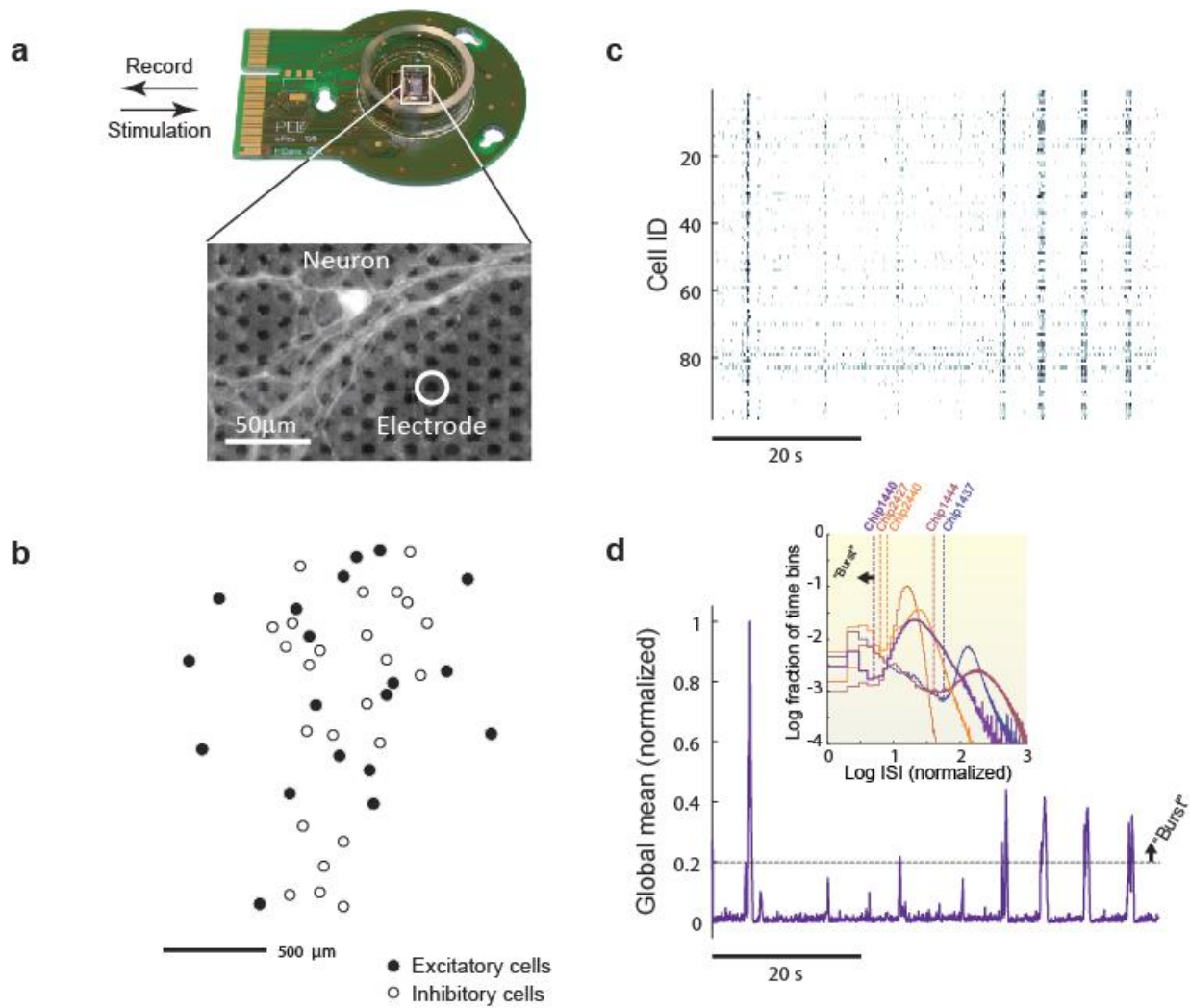

## Figure 1.

**Global bursts in cultured neural populations.**

(a) CMOS-based recording apparatus.

(b) Spatial distribution of cells recorded in a representative preparation, Chip 1440.

(c) Spontaneous activity of neurons in the same preparation, Chip 1440. The color indicates the spike count within each 5-ms time bin (white: 0 spikes; black >5 spikes). The abscissa represents the time. The majority of neurons fired together within network bursts (columns of dark-colored bins).

(d) The temporal evolution of global mean field (normalized population firing rate) computed from c. The horizontal dashed line indicates the threshold for burst detection for this preparation. (Inset) The distributions of inter-spike intervals (ISIs) in five different preparations (solid lines) and the burst detection thresholds for the individual preparations (dashed lines). The horizontal axis shows the normalized log-ISI (the logarithm of inverse global-activity; see **Methods**).

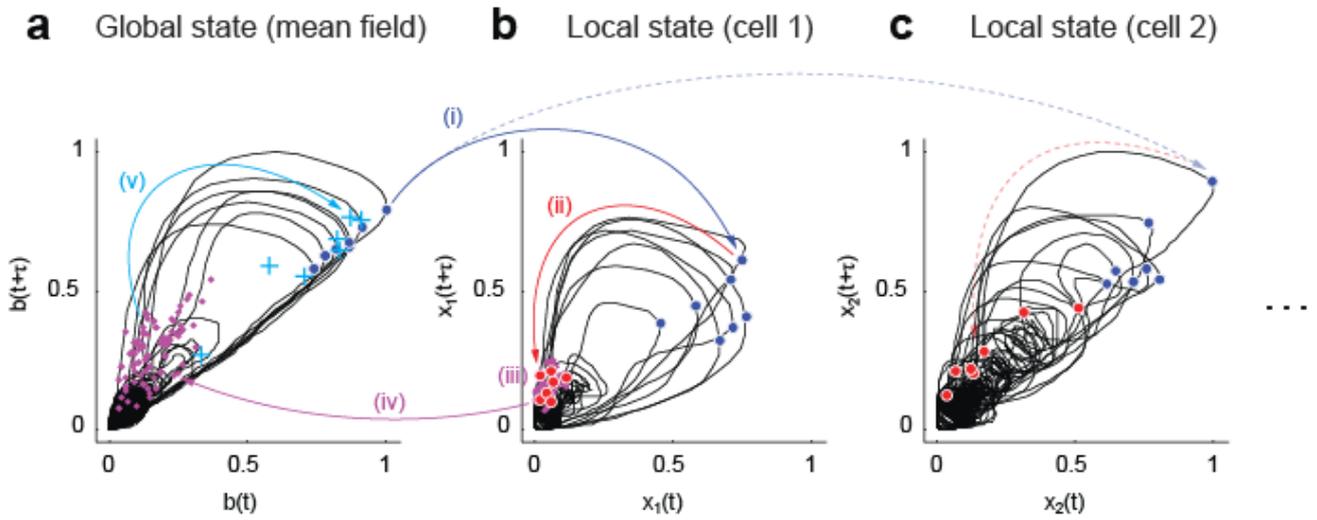

**Figure 2.**

**Trace-back embedding (TBE) method in delay coordinates relates the global population activity to the local activities.**

(a) Delay reconstruction of the mean-field dynamics ("global states").

(b, c) Delay reconstructions of single neuron activities ("local states"). Two representative neurons (Cells 1 and 2) are shown. Blue dots: the time points of peak global activities during the bursts; Red dots: the temporally preceding states from the individual global activity peaks (here, shift time = -100 ms); magenta dots: the neighboring states within the reconstructed state space for Cell 1 (10 nearest neighbors for each burst peak); cyan pluses: predicted states based on TBE method. The protocol of TBE is as follows: (i) selecting a time point $t$ on a local state trajectory (say, of neuron $i$) that corresponds to a peak population activity found in the global state trajectory (which is an *target bursting state* to be predicted), (ii) tracing-back the local state trajectory for a given time span, $\Delta t$, (iii) collecting "neighbor" time-points corresponding to states nearby $t$-$\Delta t$ in the local state trajectory (avoiding temporally too near points), (iv) mapping the neighbor time-points back into the global state trajectory, and (v) proceeding the time with $\Delta t$ on the global state trajectory to examine whether the finally obtained states (cyan pluses) cluster within bursting state and how accurately they replicate the target bursting state. The accuracy of burst prediction is higher when the traced-back states (red dotes) deviates more from the ones in non-bursting period in the local trajectory. Note that the tracing time in the retrograde direction with $\Delta t$ (step ii) and mutual mapping between global and local states (steps i and iv) are newly introduced features in this study, which extends the previous methods (24–26).

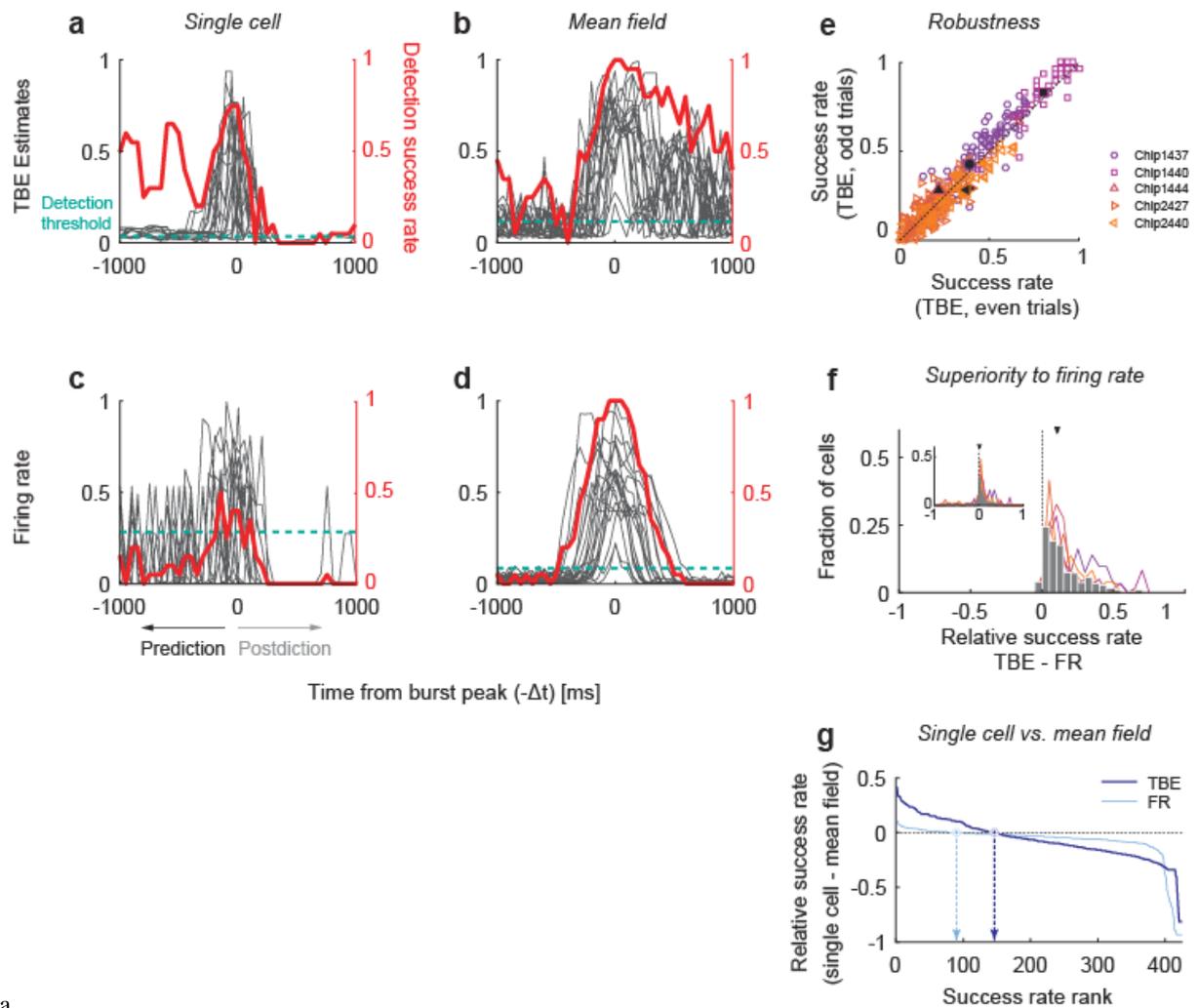

**Figure 3.**

**Presages of global burst within single neurons.**

(a) Burst detection with the TBE method based on a single representative neuron. The black traces represent global states in individual trials around burst peaks. The blue curve is the burst detection threshold for each detection time-span, $\Delta t$; global state exceeding this threshold provides burst predictor (or hallmarks of bursts after their occurrence). The threshold (dashed green line) was defined by 95 percentile of global activity outside bursting period. The red curve shows the fraction ("hit rate") of successfully detected bursts in all the burst trials.

(b) Burst detection with TBE method based on the mean-field activity in the population.

(c) Burst detection with the momentary firing rate of the same representative single neuron as the one shown in panel a.

(d) Burst detection with the momentary mean-field firing-rate.

(e) Burst predictability is a robust property of individual neurons. Dividing the data into two halves (the odd and even trials of spontaneous burst) shows that variability in burst-detection success rate is highly consistent over time. Each marker corresponds to a neuron. The filled markers represent the results using the mean field activity. The success rate was averaged over the range of time span -200 ms < $\Delta t$ < -50 ms.

(f) Burst detection with TBE method is more accurate than that based on the momentary firing rate. The comparison of success rates in burst detection with the TBE method and with the firing-rate-based method. Inset: the control analysis in which we compared the success rates of the TBE method with that of the firing-rate-based method using multiple time bins, by matching the number of bins to that used in the TBE method. The gray bars show the fraction of cells in all the preparations. The arrowheads indicate the medians. The colored lines show the fractions in the individual preparations.

(g) Single cell can outperform the mean field at predicting the spontaneous network bursts, particularly when we utilize the information in the temporal patterns of neural activity rather than the momentary activity. The light and dark blue curves respectively show the TBE- and firing-rate-based success rates in burst detection with single neurons relative to that with the mean field. For each method, the cells were sorted based on the success rate.

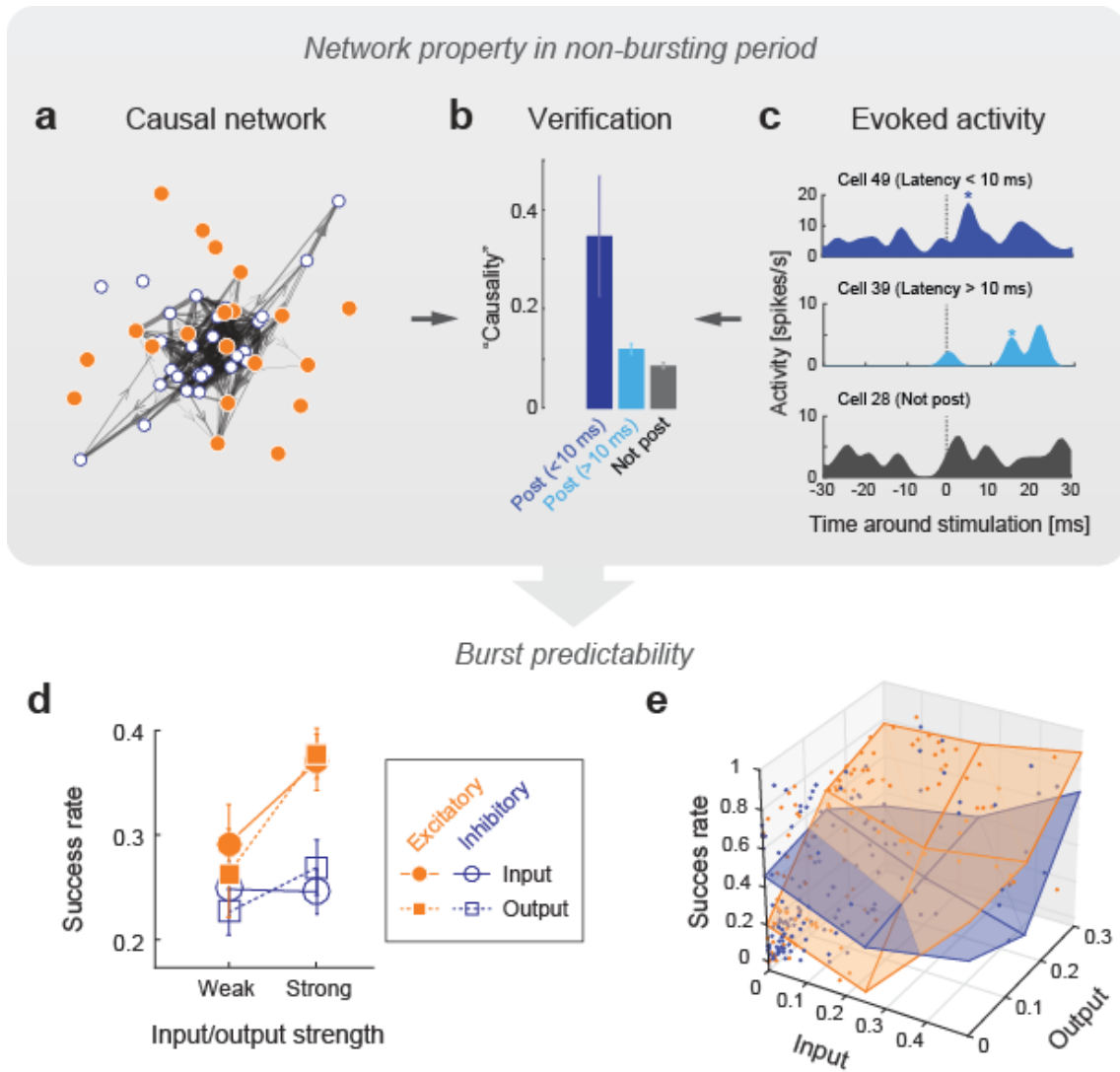

**Figure 4.**

**The network structures in non-bursting periods explains the local burst predictability.**

(a) Causal network analysis. The network depicts the directed pairwise interactions among neurons. The filled (orange) and open (blue) markers represents the putative excitatory and inhibitory cells, respectively. The thickness of arrows indicates the absolute strength of causal coupling ("causality"). For clarity of illustration, only the top 5% strong couplings are shown, and the nodes are distributed by multi-dimensional scaling with the distances defined to be inversely proportional to the causality. The figure shows the result for Chip 1440.

(b) Synaptically connected neuron pairs show larger causal interaction. The bar labels show the time span to be used for the causality analysis. Bars in dark and light blue: average cross-embedding values for cells that showed short-latent (dark, latency < 10 ms) and long (dark, latency > 10 ms) stimulus-triggered spike increase. The average causality index within time span [-100, 100] ms dissociated the post-synaptic cells from other cells (P<0.005, Wilcoxon sign rank test, independent samples). The error bar shows the s.e.m across neurons.

(c) Identification of synaptic connectivity by electrical stimulation experiment. The figure shows the peri-stimulus time histograms of three representative neurons (from top to bottom, putative direct, indirect, and non-post cells, respectively). The asterisks indicate the earliest significant increase (P>0.05) of spike count, compared to the spike count distribution in the pre-stimulus period (from -500 to -100 ms). The figure shows the spike histogram smoothed with a boxcar kernel of 5-ms width for visualization purpose.

(d) The relationship between burst predictability and the interaction strength in different neuron types (putative excitatory and inhibitory neurons). The neurons were classified into two groups depending on the strength of input or output causal couplings: the top half of neurons was labeled as "strong" whereas the bottom half was labeled as "weak." The error bar shows the s.e.m across neurons.

(e) The same as panel d, but shows the individual neurons (the dots) with a 3D plot. The surfaces represent the fitted piecewise linear functions for each neuron types. The color conventions follow that of panel d.

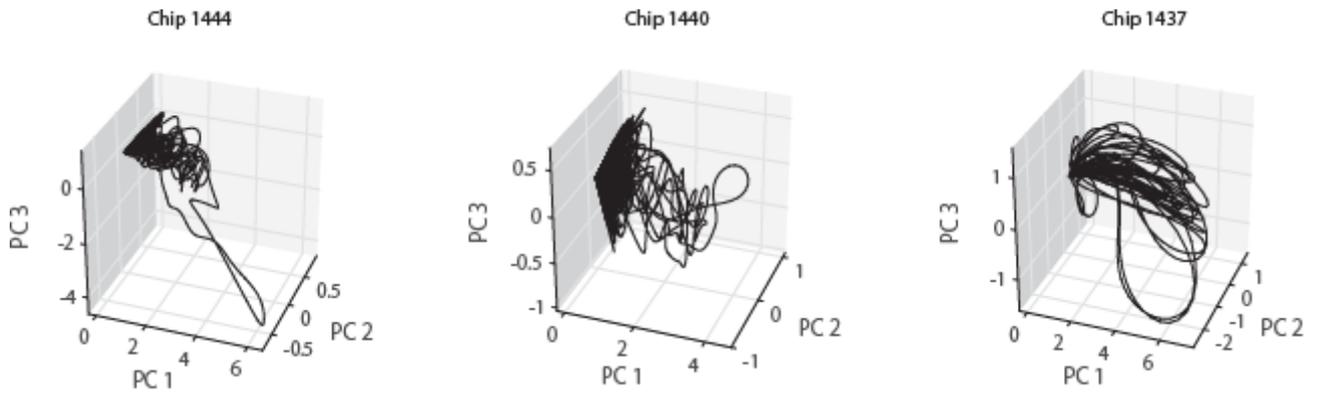

**Supplementary Figure S1.**

**The population activity dynamics visualized with principal component analysis (PCA).**

The trajectories of population activity were plotted within the space of top three principal components (PCs 1–3), by applying PCA to the multi-neuron time series for each preparation. The results for three representative preparations are shown.

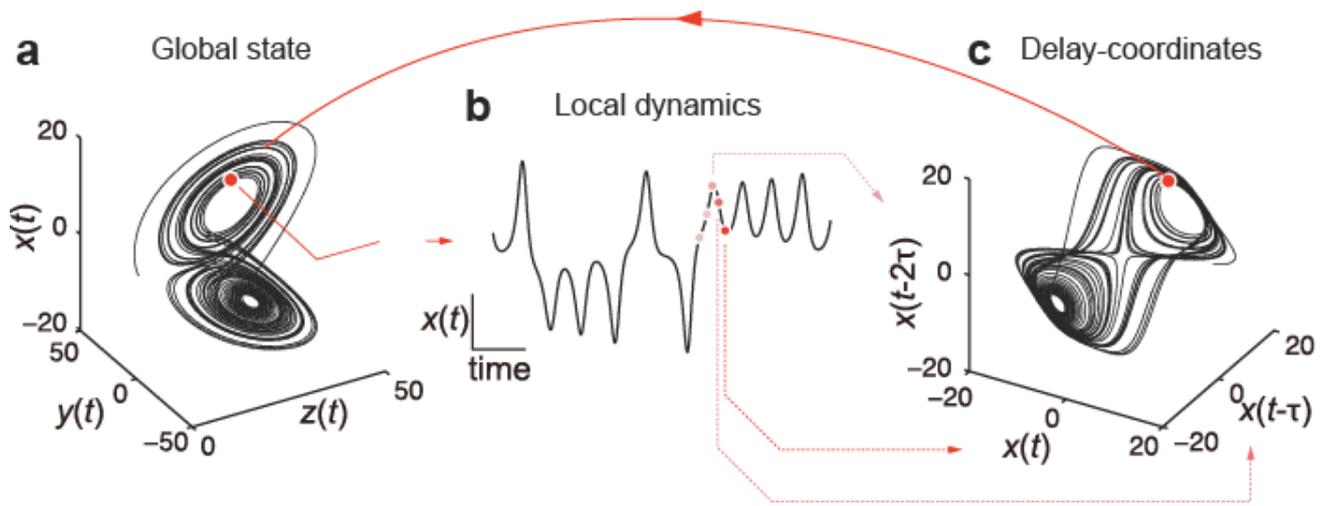

**Supplementary Figure S2.**

**The delay-embedding theorem (known as Takens' theorem).**

Suppose that we have the time-series of three variables, $(x(t), y(t), z(t))$, generated by some differential equations: $\dot{x}(t) = f(x(t), y(t), z(t))$, $\dot{y}(t) = g(x(t), y(t), z(t))$ and $\dot{z}(t) = h(x(t), y(t), z(t))$.

(a) The evolution of the global state $(x(t), y(t), z(t))$ is represented by a trajectory in the space of those three variables.

(b) Consider observing the temporal sequence of a single variable, e.g., $x(t)$.

(c) The attractor topology in the original global state space is fully recovered in a delay-coordinate of the observed variable, $(x(t), x(t-\tau), x(t-\tau))$ with an arbitrary unit delay, $\tau$. Namely, we can construct a smooth one-to-one map from the reconstructed attractor to the original attractor.

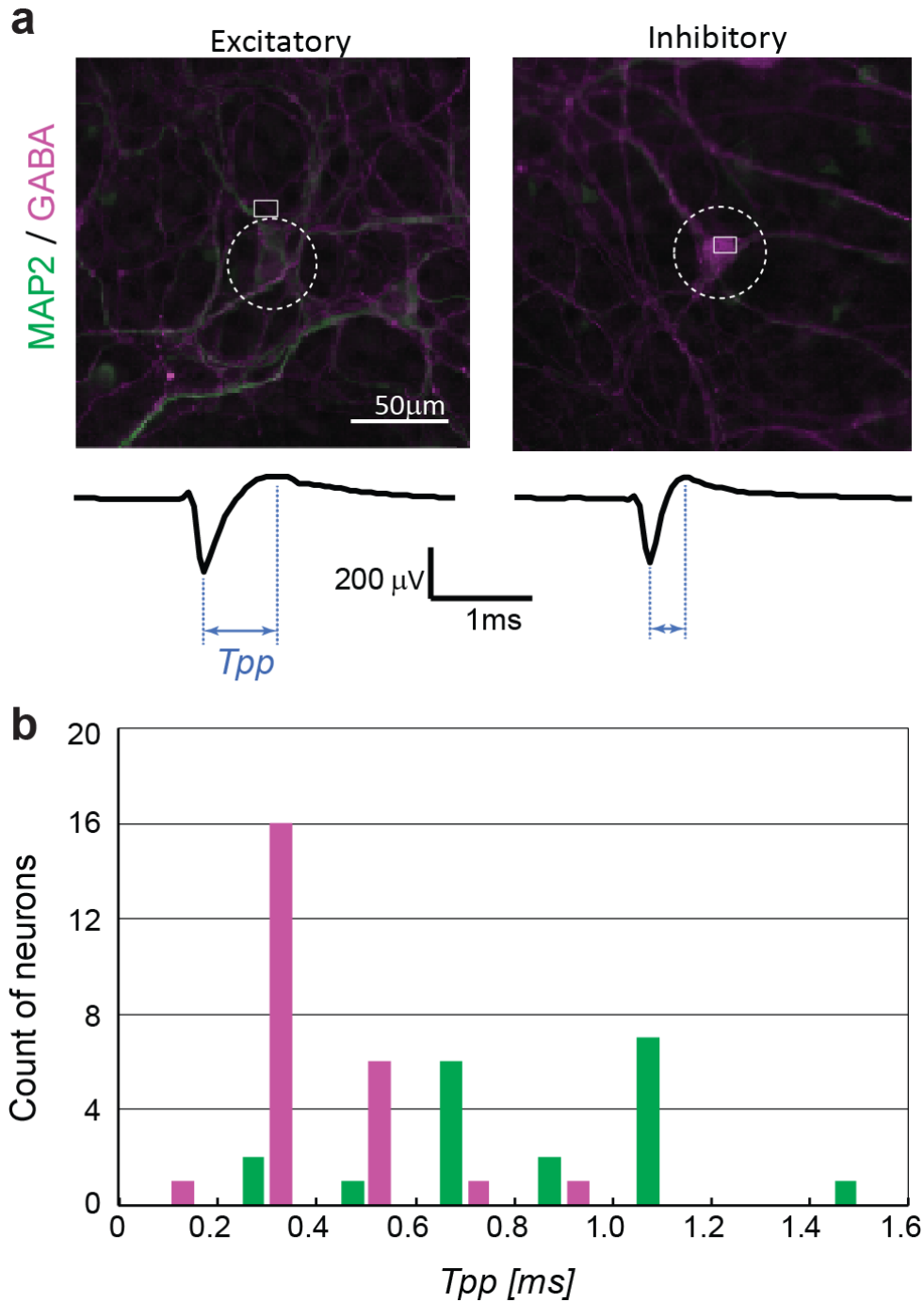

**Supplementary Figure S3.**

**Identification of Excitatory and Inhibitory Neurons**

(a) Representative neurons in immunostaining with MAP2 and GABA. Action potentials below the insets were obtained at white rectangles, putatively from a neighboring neuron in a circle. The peak-to-peak time, *Tpp*, was defined as time duration from negative peak to positive peak of action potential.

(b) Histogram of *Tpp*. Excitatory neurons (green) had larger *Tpp* than inhibitory neurons (magenta). K-means method to *Tpp* was employed to separate excitatory and inhibitory neurons.

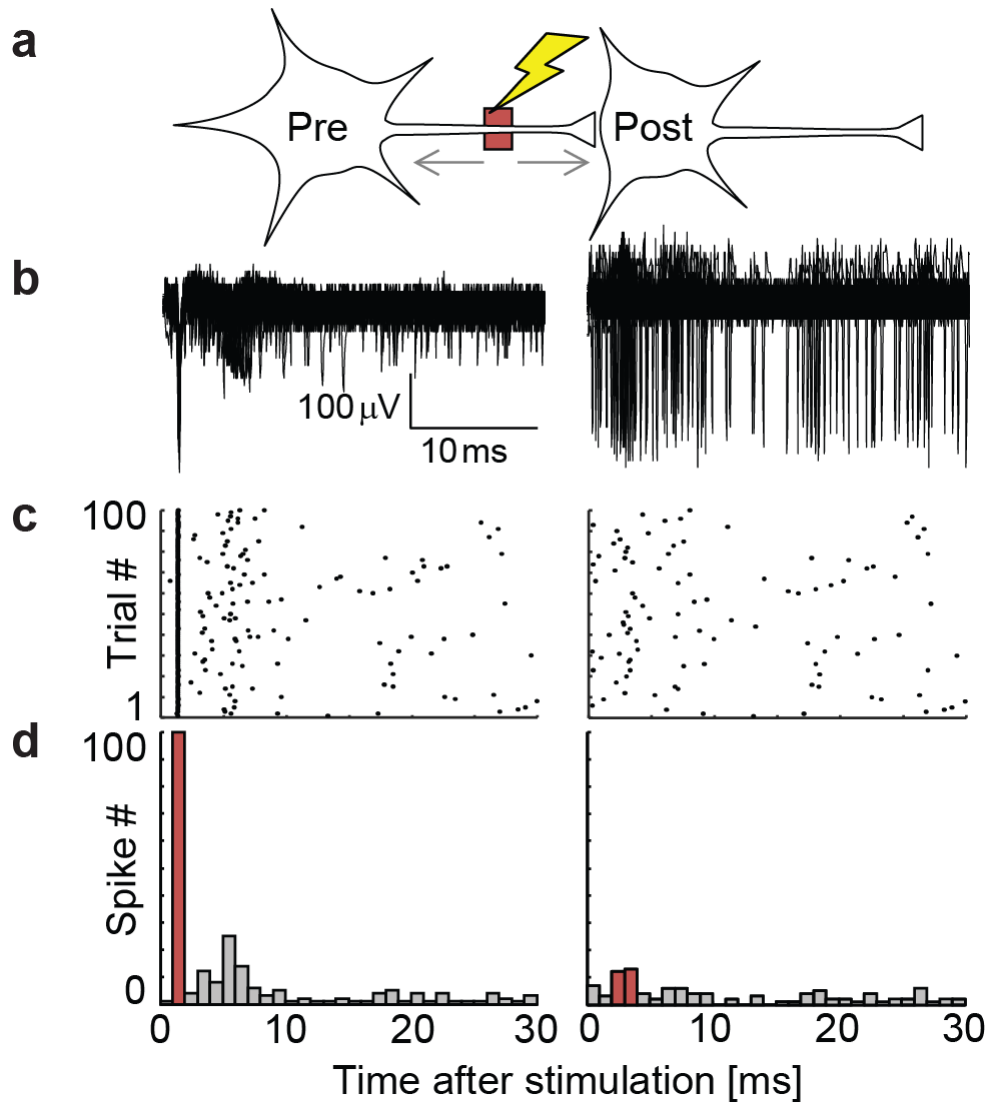

**Supplementary Figure S4.**

**Microstimulation-based estimation of synaptic connectivity in a pair-wise manner.**

(a) Axonal stimulation on an arbitrary neuron elicited bidirectional action potential propagation.

(b) Raw data of neural responses at a putative pre-synaptic neuron and post-synaptic neuron. Data from one hundred trials are superimposed.

(c) Raster plot of (b). Antidromic, direct action potentials exhibited precise temporal response, while orthodromic, synaptic action potentials were elicited stochastically with significant temporal jitters.

(d) Post-stimulus spike histograms of (c)